\documentclass[pre,twocolumn,aps,amsmath,showpacs,amsfonts,amssymb,10pt]{revtex4-1}
\usepackage{graphicx}
\usepackage{times}
\usepackage[utf8]{inputenc}
\usepackage[english]{babel}
\usepackage{fancyhdr}
\usepackage{marvosym}
\usepackage{hyperref}
\usepackage{amsmath}
\usepackage{mathrsfs}
\usepackage{latexsym}
\usepackage{amssymb}
\usepackage{graphicx} % Paquete que permite la incorporación de gráficas al documento
\usepackage[dvips,dvipsnames,usenames]{color} % Paquete que permite el uso de letras con colores
\usepackage{float} % Paquete que permite la orden [H] = Pon esta figura aquí bajo mi responsabilidad por muy feo que quede

\newcommand{\abs}[1]{\vert #1\vert}
\newcommand{\Det}[1]{\left\vert #1\right\vert}
\begin{document}

\title{Casimir Energy and Entropy in the Sphere--Sphere Geometry}
\author{Pablo Rodriguez-Lopez}
\affiliation{Departamento de F\'\i sica Aplicada I and GISC,
Facultad de Ciencias F\'\i sicas, Universidad Complutense, 28040 Madrid, Spain}

\begin{abstract}
We calculate the Casimir energy and entropy for two spheres described by the perfect metal model, plasma model, and Drude model in the large separation limit. We obtain nonmonotonic behavior of the Helmholtz free energy with separation and temperature for the perfect metal and plasma models, leading to parameter ranges with negative entropy, and also nonmonotonic behavior of the entropy with temperature and the separation between the spheres. This nonmonotonic behavior has not been found for Drude model. The appearance of this anomalous behavior of the entropy is discussed as well as its thermodynamic consequences.
\end{abstract}

\pacs{}%{42.50.Lc, 85.85.+j}

% 42.50.Lc   Quantum fluctuations, quantum noise, and quantum jumps
% 45.70.-n   Granular systems, classical mechanics of
% 45.70.Mg   Mixing granular systems
% 85.85.+j   Nanotechnology nanoelectromechanical systems, 
% 42.50.Lc   quantum noise,
% 05.40.Ca   Noise: fluctuation phenomena,

\maketitle

\section{Introduction}
In 1948, Casimir predicted the attraction between perfect metal parallel plates \cite{Casimir Placas Paralelas} and between neutral polarizable atoms \cite{VdW int. electrica} due to quantum fluctuations of the electromagnetic field. Some years later, Schwinger extended this formalism to dielectric plates at finite temperature \cite{Schwinger}. Recently, a multiscattering formalism of the Casimir effect for the electromagnetic field has been presented \cite{EGJK,Review-Jamal-Emig} (see also \cite{Multiscattering_Lambrecht} and \cite{Multiscattering_Milton}).

The Casimir effect has some peculiarities. In particular, it is a non-pairwise interaction; the Casimir thermal force (the thermal part of the Casimir energy) between two isolating bodies is not necessarily monotonic in their separation, as seen in the sphere--plate and cylinder--plate cases \cite{Metodo_Caminos_Esfera_Placa_Escalar}. In addition, for some geometries, intervals of negative entropy appear, as in the case of two parallel plates described by the Drude model \cite{Entropias_negativas_placas_Drude} or, as recently shown, in the interaction between a Drude model plate and sphere \cite{Bordag_Entropia_Esfera_Placa} and in the interaction between a perfect metal plate and sphere \cite{Canaguier-Durand_Caso_Esfera_Placa_PRA}.

In this article we study the Casimir effect between two spheres in the large separation approximation using different models for the electric permeability: (1)~the perfect metal model, (2)~the plasma model, and (3)~the Drude model. As a result, we find negative entropies in certain ranges of temperature and separation between the spheres for the perfect metal model, and for low penetration length for the plasma model. In addition, we find nonmonotonic behavior of the entropy with the separation while the force is attractive for all separations, making it appear as though the natural evolution of the system tends to increase the entropy in certain ranges of temperature and separation. For long plasma length and for the Drude model we do not find this anomalous behavior of the entropy. We also discuss the thermodynamical meaning and consequences of negative entropies in Casimir effect.

The remainder of the article is arranged as follows: In Sect. \ref{Presentacion del modelo de Emig del Efecto Casimir}, we describe the multiscattering model used herein to obtain the Casimir energies and entropies for the two spheres. In Sect. \ref{Perfect metal spheres}, we obtain the large separation limit of the Casimir energy between perfect metal spheres, as well as the entropy and force. We also obtain entropies at smaller separations numerically.
We study the plasma model system in Sect. \ref{Plasma Model} and the Drude model in Sect. \ref{Drude Model}. We discuss the thermodynamic consequences of these results in Sect. \ref{Sect:Thermodynamic}. Finally, we discuss the results obtained in the Conclusions.

%\textbf{PFA para este caso}

%\textbf{Nonmonotonous behaviour of the Casimir force have been observed in \cite{Metodo_Caminos_Esfera_Placa_Escalar} between a sphere and a plate and between a cylinder and a plate, between cylinders in presence of plates in \cite{Fuerza_cilindros_en_presencia_de_plato} and \cite{Fuerza_cilindros_en_presencia_de_plato2} and also between spheres in presence of a plate in \cite{Fuerza_esferas_en_presencia_de_plato}. This is the first time in our knowledge that nonmonotonous force between isolated spheres have been observed.}

\section{Multiscattering formalism of the Casimir energy}\label{Presentacion del modelo de Emig del Efecto Casimir}
To calculate the Casimir energy, entropy, and forces between two spheres, we employ the multiscattering formalism for the electromagnetic field \cite{EGJK,Review-Jamal-Emig}. This formalism relates the Casimir interaction between objects with the scattering of the field from each object. The Casimir contribution to the Helmholtz free energy at any temperature $T$ is given by
\begin{equation}\label{Energy_T_finite}
E = k_{B}T{\sum_{n=0}^{\infty}}'\log\Det{\mathbb{I} - \mathbb{N}(\kappa_{n})},
\end{equation}
where $\kappa_{n} = \frac{n}{\lambda_{T}}$ are the Matsubara frequencies and $\lambda_{T} = \frac{\hbar c}{2\pi k_{B}T}$ is the thermal wavelength. The prime indicates that the zero Matsubara frequency contribution has height of $1/2$. All the information regarding the system is described by the $\mathbb{N}$ matrix. For a system of two objects, this matrix is $\mathbb{N} = \mathbb{T}_{1}\mathbb{U}_{12}\mathbb{T}_{2}\mathbb{U}_{21}$. $\mathbb{T}_{i}$ is the T scattering matrix of the $i^{\text{\underline{th}}}$ object, which accounts for all the geometrical information and electromagnetic properties of the object. $\mathbb{U}_{ij}$ is the translation matrix of electromagnetic waves from object $i$ to object $j$, which accounts for all information regarding the relative positions between the objects of the system.

For a sphere of radius $R$ with electric and magnetic permeabilities $\epsilon$ and $\mu$, the $\mathbb{T}$ matrix is diagonal in $\left(\ell m P,\ell' m'P'\right)$ space, with elements given by $\mathbb{T}^{PP'}_{\ell m,\ell' m'} =  - \delta_{\ell\ell'}\delta_{mm'}\delta_{PP'}T^{P}_{\ell m}$, where the $T^{P}_{\ell m}$ are defined as \cite{Review-Jamal-Emig}
\begin{equation}\label{TMM_general}
\hspace{-5pt}T^{M}_{\ell m} \hspace{-2pt}=\hspace{-2pt} \frac{i_{\ell}(\kappa R)\partial_{R}(R i_{\ell}(n \kappa R)) \hspace{-2pt}-\hspace{-2pt} \mu\partial_{R}(R i_{\ell}(\kappa R))i_{\ell}(n \kappa R)}{k_{\ell}(\kappa R)\partial_{R}(R i_{\ell}(n \kappa R)) \hspace{-2pt}-\hspace{-2pt} \mu\partial_{R}(R k_{\ell}(\kappa R))i_{\ell}(n \kappa R)}\hspace{-1pt},
\end{equation}
\begin{equation}\label{TEE_general}
\hspace{-5pt}T^{E}_{\ell m} \hspace{-2pt}=\hspace{-2pt} \frac{i_{\ell}(\kappa R)\partial_{R}(R i_{\ell}(n \kappa R)) \hspace{-2pt}-\hspace{-2pt} \epsilon\partial_{R}(R i_{\ell}(\kappa R))i_{\ell}(n \kappa R)}{k_{\ell}(\kappa R)\partial_{R}(R i_{\ell}(n \kappa R)) \hspace{-2pt}-\hspace{-2pt} \epsilon\partial_{R}(R k_{\ell}(\kappa R))i_{\ell}(n \kappa R)}\hspace{-1pt},
\end{equation}
where $i_{\ell}(x) = \sqrt{\frac{\pi}{2x}}I_{\ell + \frac{1}{2}}(x)$, $k_{\ell}(x) = \frac{2}{\pi}\sqrt{\frac{\pi}{2x}}K_{\ell + \frac{1}{2}}(x)$, and $n = \sqrt{\mu\epsilon}$. Expressions for the $\mathbb{U}_{\alpha\beta}$ matrices can be found in \cite{Review-Jamal-Emig} and \cite{Wittmann}.

\section{Perfect metal spheres}\label{Perfect metal spheres}
In the perfect metal limit, we apply $\epsilon\to\infty$ for any $\mu$ to Eqs. \eqref{TMM_general} and \eqref{TEE_general}. In this case, the $\mathbb{T}$ matrix elements are independent of $\mu$, taking the well-known universal form
\begin{equation}
\mathbb{T}^{MM}_{\ell m,\ell' m'} = - \delta_{\ell\ell'}\delta_{mm'}\frac{\pi}{2}\frac{I_{\ell + \frac{1}{2}}(\kappa R)}{K_{\ell + \frac{1}{2}}(\kappa R)},
\end{equation}
\begin{equation}
\mathbb{T}^{EE}_{\ell m,\ell' m'} \hspace{-2pt}=\hspace{-2pt} - \delta_{\ell\ell'}\delta_{mm'}\frac{\pi}{2}\frac{\ell I_{\ell + \frac{1}{2}}(\kappa R) - \kappa R I_{\ell - \frac{1}{2}}(\kappa R)}{\ell K_{\ell+ \frac{1}{2}}(\kappa R) + \kappa R K_{\ell - \frac{1}{2}}(\kappa R)}.
\end{equation}

\subsection{Casimir energy in the large separation limit}
\begin{figure}[H]
\begin{center}
\includegraphics{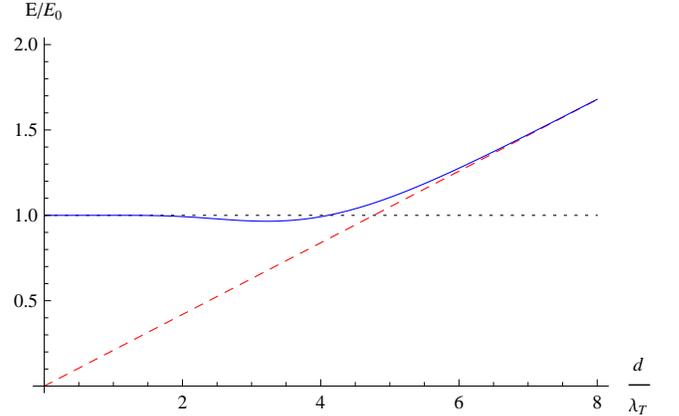}
\caption{(Color online) Casimir energy between perfect metal spheres as a function of $\frac{d}{\lambda_{T}}$ compared with the energy at zero temperature. The dotted curve is the quantum limit, the dashed curve is the classical limit, and the solid curve is the asymptotic finite-temperature Casimir energy. Note that these curves are independent of the radius of the spheres but are only valid in the large separation limit.}
\label{Energia_2_esferas_metalicas}
\end{center}
\end{figure}

To obtain the large separation limit of the Casimir energy, we need the dominant part of the $\mathbb{T}$ matrix in this limit. We define the adimensional frequency $q$ by $\kappa = q/d$, where $d$ is the distance between the centre of the spheres. The main contribution in the large separation limit comes from the lowest-order expansion term of the $\mathbb{T}$ matrix elements in $1/d$. As it is known that, at small $\kappa$, the $\mathbb{T}$ elements scale as $\kappa^{2\ell+1}$ \cite{EGJK}, the dominant contribution comes from the dipolar polarizabilities part of the $\mathbb{T}$ matrix, taking the form
\begin{equation}
\mathbb{T}^{MM}_{1 m,1 m'} = - \frac{1}{3}\left(\frac{qR}{d}\right)^{3},
\hspace{20pt}
\mathbb{T}^{EE}_{1 m,1 m'} =   \frac{2}{3}\left(\frac{qR}{d}\right)^{3}.
\end{equation}
By the use of the universal relationship between determinants and traces, $\log\abs{A} = \text{Tr}\log(A)$, and applying a Taylor expansion in terms of $\frac{1}{d}$ of Eq. \eqref{Energy_T_finite}, we obtain
\begin{equation}
E = - k_{B}T{\sum_{n = 0}^{\infty}}'\text{Tr}\left(\mathbb{N}(\lambda_{T}^{-1}n)\right).
\end{equation}
Using the translation matrices in a spherical vector multipole basis \cite{Review-Jamal-Emig} and the large separation approximation of the $\mathbb{T}$ matrix, the trace of the $\mathbb{N}$ matrix is obtained by straightforward calculus.
Here we denote with a sub--index $T$ the results valid for all temperatures, with a sub--index $0$ the results in the quantum limit ($T\to 0$), and with a sub--index $cl$ the results in the classical limit ($\hbar\to 0$), which is equivalent to the high $T$ limit.
Carrying out the sum over Matsubara frequency, we obtain the Casimir contribution to the Helmholtz free energy as
\begin{eqnarray}\label{Energia_finite_T_metal_perfecto}
E_{T} & = & - \frac{\hbar c R^{6}}{2\pi d^{7}}\frac{z e^{5z}}{2\left(e^{2z} - 1\right)^{5}}\times\nonumber\\
& & \left(2\left(15 - 29z^{2} + 99z^{4}\right)\cosh(z) + 15\cosh(5z)\right.\nonumber\\
& & \left. + \left( - 45 + 58z^{2} + 18z^{4}\right)\cosh(3z)\right.\\
& & \left. + 24z\left( 6z^{2} - 5 + \left(5 + 3z^{2}\right)\cosh(2z)\right)\sinh(z)\right),\nonumber
\end{eqnarray}
where $z = d/\lambda_{T}$. We also define the adimensional Casimir energy as $E_{ad}(z) = \frac{2\pi d^{7}}{\hbar c R^{6}}E_{T}$. From this result, the quantum ($T\to 0$) and classical ($\hbar\to 0$) limits with their first corrections are easily obtained as
\begin{equation}\label{Low_T_corrections_Energy}
E_{0} = - \frac{143 \hbar c R^{6}}{16\pi d^{7} } - \frac{8 k_{B}T\pi^{5}R^{6}}{27 d}\left(\frac{k_{B}T}{\hbar c}\right)^{5},
\end{equation}
\begin{eqnarray}\label{High_T_corrections_Energy}
E_{cl} & = & - \frac{15 k_{B}T R^{6}}{4 d^{6}}\\
& & - k_{B}T\frac{R^{6}}{2d^{6}}\left(15 + 30 z + 29 z^{2} + 18 z^{3} + 9 z^{4}\right)e^{-2z},\nonumber
\end{eqnarray}
where $z = d/\lambda_{T}$. Note that in Eq. \eqref{Low_T_corrections_Energy}, the first correction to zero temperature case is proportional to $T^{6}$, contrary to the plate--sphere case \cite{Canaguier-Durand_Caso_Esfera_Placa_PRA, Metodo_Caminos_Esfera_Placa_Escalar} and to the cylinder--plate case \cite{Metodo_Caminos_Esfera_Placa_Escalar}, where a result proportional to $T^{4}$ were obtained for both systems. The result presented here is new for perfect metal spheres, but expected, because it has been obtained also for compact objects in the diluted limit \cite{PSA_EM_general, Barton}. As corrections have the same negative sign of the Casimir energy in both limits, they describe an increase of the magnitude of the Casimir energy in high and low temperature limits. It is no longer the case if we study the energy for all temperatures. In fact, for some ranges of separation and temperature, thermal photons tend to reduce the Casimir energy between the spheres, as shown in Fig. \ref{Energia_2_esferas_metalicas}. This effect is captured in the contribution of the next term to the quantum limit, this term is $\Delta E_{0} = +\frac{2288 d k_{B}T\pi^{7}R^{6}}{1575}\left(\frac{k_{B}T}{\hbar c}\right)^{7}$ and tends to reduce the magnitude of the energy. Fig. \ref{Energia_2_esferas_metalicas} is an indicator of the appearance of negative entropy in this system because of the negative slope of the energy curve.
It is clear in Fig. \ref{Energia_2_esferas_metalicas} that for some temperatures and distances, the Casimir energy between spheres is lower than in the zero temperature case. But it is less evident the validity Eq. \eqref{Low_T_corrections_Energy}. In fact, there is a tiny increase of $E/E_{0}$ at low $\frac{d}{\lambda_{T}}$ until reaching a local maximun at $\frac{d}{\lambda_{T}}\approx 1.0388$ of $E/E_{0}\approx 1 + 10^{-4}$. It is not visible in Fig. \ref{Energia_2_esferas_metalicas} because the difference os scales. The local minimum is reached at $\frac{d}{\lambda_{T}}\approx 3.21733$, where $E/E_{0}\approx 1 - 3.46\times 10^{-2}$. This minimum is reached, in the important cases of room temperature ($T = 300K$) and at the boiling point of $\text{N}_{\text{2}}$ ($T = 77K$) at $d_{T=300} = 3.89\mu m$ and $d_{T=77} = 15.21\mu m$ respectively.

\subsection{Casimir entropy in the large separation limit}
\begin{figure}[H]
\begin{center}
\includegraphics{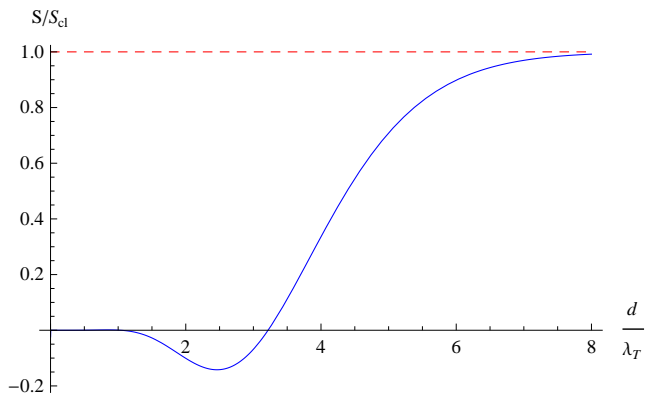}
\caption{(Color online) Casimir entropy between perfect metal spheres as a function of $\frac{d}{\lambda_{T}}$ compared with the classical limit. The dashed curve is the classical limit, and the solid curve is the asymptotic finite-temperature Casimir entropy. Note the regions of negative entropy and negative slope of the entropy with the parameter $\frac{d}{\lambda_{T}}$}
\label{Entropia_2_esferas_metalicas}
\end{center}
\end{figure}

In the canonical ensemble, the entropy is defined as $S = - \partial_{T}E$, where we remain that we denote the Helmholtz free energy by $E$. In the large separation limit, the Helmholtz free energy depends on the adimensional variable $z = \frac{d}{\lambda_{T}} = 2\pi\frac{dk_{B}T}{\hbar c}$, so we can write the entropy as
\begin{equation}
S = - \frac{\partial z}{\partial T}\frac{\partial E}{\partial z} = - 2\pi\frac{dk_{B}}{\hbar c}\frac{\partial E}{\partial z},
\end{equation}
and define the adimensional entropy as $S_{ad}(z) = \frac{d^{6}}{k_{B}R^{6}}S = - \partial_{z}E_{ad}(z)$. From this result, the quantum ($T\to 0$) and classical ($\hbar\to 0$) limits are easily obtained as
\begin{equation}\label{Low_T_Entropy}
S_{0} = 0 + \frac{16  k_{B} \pi^{5}R^{6} }{9 d }\left(\frac{k_{B}T}{\hbar c}\right)^{5},
\end{equation}
\begin{eqnarray}\label{High_T_Entropy}
S_{cl} & = & \frac{15 k_{B} R^6}{4 d^6}\\
& & + k_{B}\frac{R^{6}}{2d^{6}}\hspace{-1pt}\left(15 \hspace{-1pt}+\hspace{-1pt} 30z \hspace{-1pt}+\hspace{-1pt} 27z^{2} \hspace{-1pt}+\hspace{-1pt} 14z^{3} \hspace{-1pt}+\hspace{-1pt} 9z^{4} \hspace{-1pt}-\hspace{-1pt} 18z^{5}\right)\hspace{-1pt}e^{-2z}\hspace{-2pt},\nonumber
\end{eqnarray}
where $z = \frac{d}{\lambda_{T}}$. So, the entropy is a growing function with temperature in both limits, but this is not the case for all temperatures (the next low temperature expansion term of Eq. \eqref{Low_T_Entropy} is $\Delta S_{0} = -\frac{18304\pi^{7}}{1575}k_{B}d R^{6}\left(\frac{k_{B}T}{\hbar c}\right)^{7}$, which already indicates that the entropy could change its growing behavior with the temperature), as we can observe in Fig. \ref{Entropia_2_esferas_metalicas}, where a region of negative entropy and another region of negative slope of the entropy are observed.

Because of the limit at low temperature of the entropy, we know that the entropy is positive for low $T$ (not seen in Fig. \ref{Entropia_2_esferas_metalicas} because it is small compared with $S/S_{\text{cl}}$, but it can be observed in Fig. \ref{LogLogPlot_Entropia_2_esferas_metalicas_forall_distances}), so there are three points where $S=0$, including the origin. These two new points where the entropy nulls correspond to the local maximum and minimum observed in the Casimir energy in Fig. \ref{Energia_2_esferas_metalicas}.

Negative entropy of the Casimir effect has already been obtained between Drude parallel plates in \cite{Entropias_negativas_placas_Drude} and in \cite{Ingold}, and more recently between a perfect metal plate and sphere in \cite{Canaguier-Durand_Caso_Esfera_Placa_PRA} and between a Drude sphere and plate in \cite{Zandi_Emig_Placa_Esfera_varios_modelos} and \cite{Bordag_Entropia_Esfera_Placa}. This is the first time, to the best of our knowledge, that negative entropy appears between spheres because of the Casimir effect.

These results are only valid when the separation between the spheres is large compared with their radius, regardless of the radius of each one. Therefore, it is possible that the interval of negative entropy would disappear when the separation between the spheres reduces, regardless of the temperature of the system. A numerical exploration of entropies at smaller separation has been performed to verify this issue.

\subsection{Numerical study at smaller separations}
As noted in the previous subsection, asymptotic results are no longer valid when the separation between the spheres becomes comparable to their radius. For this reason, a numerical study of entropy was performed for these cases. We computed Eq. \eqref{Energy_T_finite} numerically for all temperatures from $T=0$ until reaching the classical limit for fixed ratio between the radius $R$ and separation, $r = \frac{R}{d}$.

For spheres, the $\mathbb{T}$ matrices are diagonal but infinite matrices, so a cutoff in $(\ell,\ell')$ space is needed to obtain finite-dimensional matrices. In addition, another cutoff in Matsubara frequency is needed to obtain a finite series.

The proposed method is an asymptotic approximation, at small separations more and more modes are needed to obtain convergent results. This means that there exists a minimum separation between the spheres below which we are not able to take into account enough multipoles to obtain good results. We use multipoles up to $\ell \leq 15$, which means that we are restricted to $r_{\text{max}} = \frac{R}{d_{\text{min}}} \leq 0.45$, when contact is reached at $r_{\text{contact}} = \frac{R}{d_{\text{contact}}} = \frac{R}{2R} = 0.5$, where the energy diverges.

In the small separation limit, the proximity force approximation (PFA) is known to be a good approximation to the Casimir energy \cite{EGJK}. It is also known that perfect metal plates do not experience negative entropies, so we do not expect to observe negative entropies between spheres in the small separation limit.

\begin{figure}[H]
\begin{center}
\includegraphics{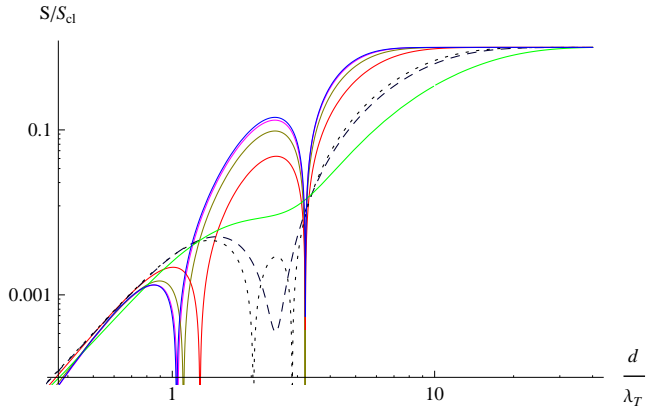}
\caption{(Color online) Log--log plot of the absolute value of the entropy divided by its classical limit for perfect metal spheres at constant $r$ as a function of $\frac{d}{\lambda_{T}}$. Starting from the asymptotic solid blue curve ($r\to 0$), we increase $r$ in steps of $0.1$ up to $r = 0.4$ (magenta curve for $r=0.1$, yellow for $r=0.2$, red for $r=0.3$ and $r=0.4$ is the dotted curve). Dashed curve is the case $r = 0.41$, where the interval of negative entropy has disappeared. The curve for $r=0.45$ (solid green curve) is also shown}
\label{LogLogPlot_Entropia_2_esferas_metalicas_forall_distances}
\end{center}
\end{figure}

In Fig. \ref{LogLogPlot_Entropia_2_esferas_metalicas_forall_distances}, the entropy of the system of two perfect metal spheres is plotted as a function of $z = \frac{d}{\lambda_{T}}$ for constant $r = \frac{R}{d}$. The large separation result is shown too. We choose a log--log representation of the absolute value of the entropy divided by its corresponding classical limit. Therefore, zeros are observed as log divergences, and we can also observe the cases of negative entropy. Starting in the large separation regime, we observe an interval of negative entropy. As we increase $r$ (reducing the separation between the spheres), the region of negative entropy tends to reduce until it disappears between $r = 0.40$ and $r = 0.41$. Power-law decay of the entropy \eqref{Low_T_Entropy} at low temperatures is observed as a linear decay of the curve at low $z$, and constant behavior in the high-temperature limit \eqref{High_T_Entropy} is also reached in the computation.

The power law decay of the entropy reduces from $S\propto T^{5}$ for asymptotic results ($r\to 0$) to around  $S\propto T^{3}$ for the closest studied case ($r = 0.45$).

\subsection{Casimir force in the large separation limit}
\begin{figure}[H]
\begin{center}
\includegraphics{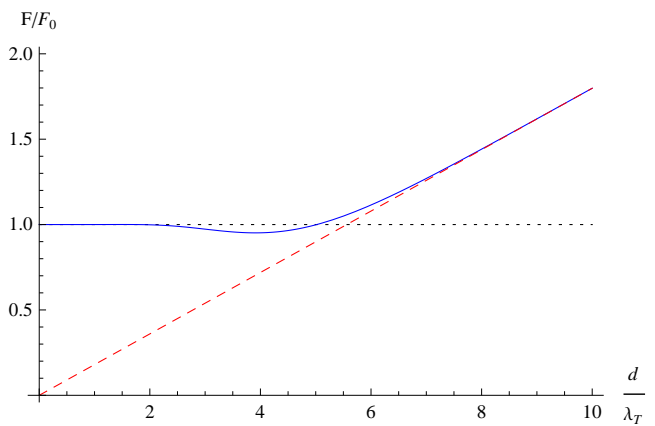}
\caption{(Color online) Asymptotic force divided by its zero-temperature limit as a function of $\frac{d}{\lambda_{T}}$. The dotted curve is the zero-temperature result. The dashed curve is the classical limit, and the solid curve is the result at finite temperatures. The nonmonotonic behavior of the force with temperature results from the negative slope of the solid curve at constant separation}
\label{Fuerza_2_esferas_metalicas}
\end{center}
\end{figure}

In the previous subsection we demonstrated that intervals of negative entropy appear due to the Casimir effect between perfect metal spheres. In these cases, for any given temperature, a minimum of entropy exists for a given, nonzero separation. Naively, this would imply a violation of the third law of thermodynamics if the Casimir force is not zero at the minimum of the entropy. However, this is not the case, as will be explained in Sect. \ref{Sect:Thermodynamic}. In addition to that, we find that the force is always attractive, independent of the increase or decrease of the entropy, which also would naively imply a violation of the second law, because the system can be enforced to perform a process in which the entropy is reduced instead increased. However, this is not the case, as will be also explained in Sect. \ref{Sect:Thermodynamic}. However, the appearance of negative entropy does have an effect on the force. The asymptotic Casimir force $F = -\partial_{d}E$ can be written in terms of the adimensional Casimir energy as
\begin{equation}\label{Fuerza_adimensional}
F_{ad}(z) = \frac{2\pi}{\hbar c}\frac{d^{8}}{R^{6}}F = 7E_{ad}(z) - z\partial_{z}E_{ad}(z),
\end{equation}
where $z = \frac{d}{\lambda_{T}}$. In Fig. \ref{Fuerza_2_esferas_metalicas}, the adimensional asymptotic force between the perfect metal spheres compared with the zero-temperature force is plotted as a function of $\frac{d}{\lambda_{T}}$. Here, for constant temperature, we observe nonmonotonic behavior of the force with the adimensional parameter $\frac{d}{\lambda_{T}}$.

Nonmonotonic behavior of the adimensional Casimir force implies nonmonotonic force behavior with temperature, but not with separation, because of the extra dependence of the force on the separation in Eq. \eqref{Fuerza_adimensional}. In fact, it is easy to verify that the force behaves monotonically with separation, and the nonmonotonicity of the entropy with separation implies nonmonotonic behavior of the force with temperature, because
\begin{equation}\label{Relacion_Fuerza_Entropia}
\frac{\partial F}{\partial T} = - \frac{\partial^{2} E}{\partial T\partial d} = \frac{\partial S}{\partial d},
\end{equation}
so the appearance of negative slopes of the entropy with separation implies nonmonotonicity of the Casimir force with temperature, despite the attractive force for all separations and temperatures.

%As observed in Fig. \ref{Fuerza_2_esferas_metalicas}, force between spheres is always atractive, then asymptotically for any given temperature, exist a range of distances where the system tends to increase the entropy because Casimir effect.

As observed in Fig. \ref{Fuerza_2_esferas_metalicas}, the force between the spheres is always attractive, but it is not monotonic with temperature; asymptotically, for any given temperature, there exists a range of separations for which the force decreases with temperature. This is the first time, to the best of our knowledge, that nonmonotonic behavior of the Casimir force with temperature has been described between compact objects. Nonmonotonicities of the force between a plate and cylinder and between a plate and sphere were already obtained in \cite{Metodo_Caminos_Esfera_Placa_Escalar}, but in that case the nonmonotonicity already appears for the scalar field. Nonmonotonicity does not appear between spheres for the scalar field; this is a characteristic effect of the electromagnetic field, because cross-polarization terms of the Casimir energy are essential for this nonmonotonicity to appear.

\section{Plasma model}\label{Plasma Model}
In this section, we assume that the electric susceptibility of both spheres is described by the plasma model, i.e.,
\begin{equation}
\epsilon(ic\kappa) = 1 + \frac{(2\pi)^{2}}{\lambda_{P}^{2}\kappa^{2}}
\end{equation}
and $\mu = 1$. To obtain the large separation limit of the Casimir energy, we need the dominant part of the $\mathbb{T}$ matrix in this limit. The main contribution comes from the dipolar polarizabilities part of $\mathbb{T}$ matrix, taking the form \cite{Canaguier-Durand_Caso_Esfera_Placa_PRA,Zandi_Emig_Placa_Esfera_varios_modelos}
\begin{equation}
\mathbb{T}^{MM}_{1 m,1 m'} = - \frac{R\lambda_{P}^{2}q^{3}}{12\pi^{2}d^{3}} \left(3 + y^{2} - 3y\coth\left(y\right)  \right),
\end{equation}
\begin{equation}
\mathbb{T}^{EE}_{1 m,1 m'} =   \frac{2}{3}\left(\frac{qR}{d}\right)^{3},
\end{equation}
with $y = 2\pi\frac{R}{\lambda_{P}}$. The coefficients of the magnetic polarizability depend on the plasma frequency, tending to the perfect metal result as $\lambda_{P}\to 0$ and to zero in the transparent limit $\lambda_{P}\to\infty$ \cite{Zandi_Emig_Placa_Esfera_varios_modelos}.

\subsection{Casimir energy in the large separation limit}
Once we have the asymptotic $\mathbb{T}$ matrix, the Casimir energy can be obtained by a straightforward but long calculation. It is possible to obtain analytical results for finite temperatures, but they are too long to show here. The zero-temperature Casimir energy is
\begin{eqnarray}
E_{0} & = & - \frac{\hbar c R^{6} }{16\pi d^{7}y^{4}}\times\\
& & \left( 207 + 222 y^{2} + 143 y^{4} - 414y\coth(y)\right.\nonumber\\
& & \left. + 207 y^{2}\coth(y)^{2} - 222 y^{3}\coth(y)\right),\nonumber
\end{eqnarray}
and the high-temperature limit is given by
\begin{equation}
E_{cl} \hspace{-2pt}=\hspace{-2pt} - k_{B}T\frac{3 R^{6}}{d^{6}}\left(1 \hspace{-2pt}+\hspace{-2pt} \frac{1}{4y^{4}}\left(3 \hspace{-2pt}+\hspace{-2pt} y^{2} \hspace{-2pt}-\hspace{-2pt} 3y\coth(y)\right)^{2}\right),
\end{equation}
with $y = 2\pi\frac{R}{\lambda_{P}}$. The perfect metal limit is reached for $\lambda_{P}\to 0$, as expected. When the plasma wavelength $\lambda_{P} \gg R$, only the electric sector of the $\mathbb{T}$ matrix contributes, so in this case the energy in the zero- and high-temperature limits is given by
\begin{equation}
E_{0} = -\frac{23 \hbar c R^{6} }{4\pi d^{7}},
\hspace{20pt}
E_{cl} = - k_{B}T\frac{3 R^{6}}{d^{6}}.
\end{equation}

\subsection{Casimir entropy in the large separation limit}
\begin{figure}[h]
\begin{center}
\includegraphics{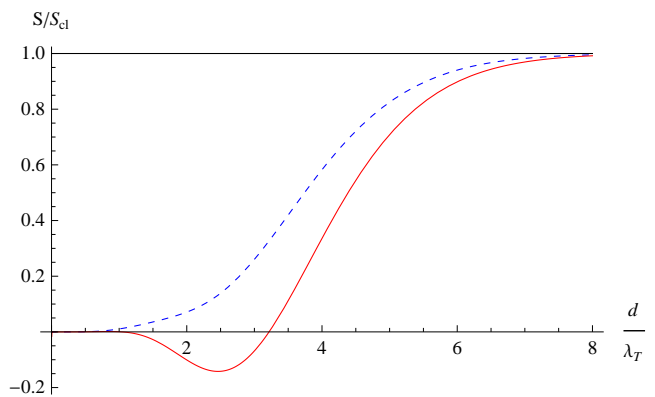}
\caption{(Color online) Entropy between plasma spheres as a function of $\frac{d}{\lambda_{T}}$ divided by the classical limit for the two limit cases. The solid curve is the entropy in the perfect metal limit $\lambda_{R} \ll R$, where a region of negative entropy and nonmonotonicities of entropy with separation and temperature are present; the dashed curve is the entropy in the transparent limit $\lambda_{R} \gg R$, where the usual monotonic behavior of entropy is shown}
\label{Entropia_2_esferas_Plasma_Casos_Limite}
\end{center}
\end{figure}

\begin{figure}[h]
\begin{center}
\includegraphics{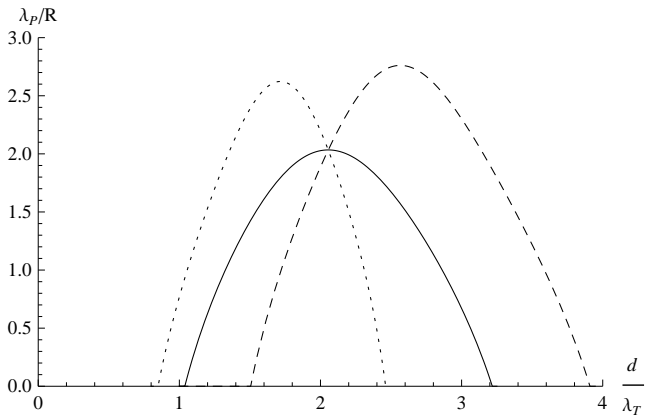}
\caption{$(\frac{d}{\lambda_{T}},\frac{\lambda_{P}}{R})$ map of zeros of the entropy (dotted curve), the temperature derivative of the entropy (dashed curve), and the separation derivative of the entropy (dotted curve)}
\label{Ceros_Entropia_2_esferas_Plasma}
\end{center}
\end{figure}

The entropy of the system is obtained in the same way as for the perfect metal case, but for the plasma model we have two different regimes. When $\lambda_{P}\ll R$, we reach the perfect metal limit, obtaining negative entropy and nonmonotonic behavior of the entropy with separation and temperature as before. However, when $\lambda_{P}\gg R$, the spheres are transparent to the magnetic field and the problem reduces to a scalar problem. In this case, the entropy is always positive. In Fig. \ref{Entropia_2_esferas_Plasma_Casos_Limite}, the adimensional asymptotic entropy, compared with its classical limit, is plotted as a function of $\frac{d}{\lambda_{T}}$ for these two limits. In Fig. \ref{Ceros_Entropia_2_esferas_Plasma}, we present the points $(\frac{d}{\lambda_{T}},\frac{\lambda_{P}}{R})$ corresponding to zeros of the entropy (continuous curve), its temperature derivative (dashed curve), or its separation derivative (dotted curve).
As discussed in the case of perfect metal plates, the anomalous behavior of the entropy found here would naively imply a violation of second and third laws of thermodynamics, because the Casimir force is always attractive, irrespective of the slope of the entropy. However, this is not the case, as we will discuss in Sect. \ref{Sect:Thermodynamic}.
As noted above, in the perfect metal limit we obtain a region of negative entropy, while for low $\lambda_{P}$ this region disappears. The point where negative entropies appear depends on the magnetic susceptibility $\mu$. For $\mu = 1$, negative entropies appear around $\lambda_{P} \approx 2 R$.

\section{Drude model}\label{Drude Model}
In this section we assume that the electric susceptibility of both spheres is described by the Drude model, i.e.,
\begin{equation}
\epsilon(ic\kappa) = 1 + \frac{4 \pi^{2}}{\lambda_{P}^{2}\kappa^{2} + \frac{\pi c\kappa}{\sigma}}
\end{equation}
and $\mu = 1$. To obtain the large separation limit of the Casimir energy, we need the dominant part of the $\mathbb{T}$ matrix in this limit. The main contribution comes from the dipolar polarizabilities part of the $\mathbb{T}$ matrix, taking the form \cite{Canaguier-Durand_Caso_Esfera_Placa_PRA,Zandi_Emig_Placa_Esfera_varios_modelos}
\begin{equation}
\hspace{-4pt}\mathbb{T}^{MM}_{1 m,1 m'} \hspace{-2pt}=\hspace{-2pt} - \frac{4\pi}{45}\frac{R\sigma}{c}\left(\frac{qR}{d}\right)^{4}\hspace{-2pt},
\hspace{10pt}
\mathbb{T}^{EE}_{1 m,1 m'} \hspace{-2pt}=\hspace{-2pt}   \frac{2}{3}\left(\frac{qR}{d}\right)^{3}\hspace{-2pt}.
\end{equation}
Now, the response of the material changes dramatically from the case of the plasma model. In principle, one would expect to obtain the $\mathbb{T}$ matrix for the plasma model as $\sigma\to\infty$, but it is obvious that this is not the case. Not only do we not recover the plasma model in this limit, but also the $\mathbb{T}$ matrix diverges. The reason is that the Plasma model is some kind of singular limit of Drude model, with a qualilative different behaviour. Now the dominant contribution is given by the electric sector of the $\mathbb{T}$ matrix, whereas the magnetic part does not contribute for asymptotic energies.

\subsection{Casimir energy and entropy in the large separation limit}
\begin{figure}[h]
\begin{center}
\includegraphics{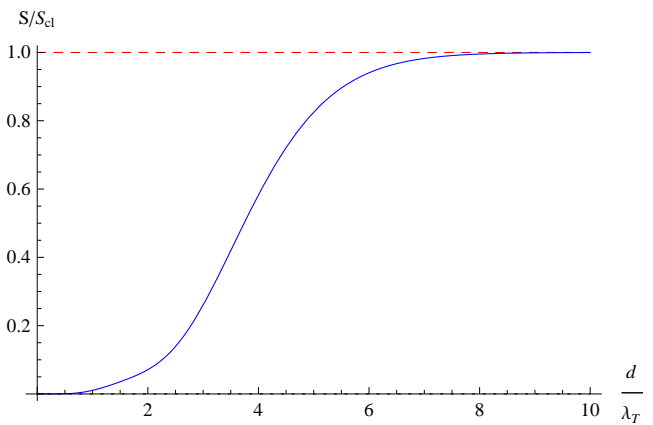}
\caption{(Color online) Entropy as a function of the adimensional parameter $\frac{d}{\lambda_{T}}$ divided by its classical limit for the large separation limit of the system of Drude spheres. Entropy is positive and monotonic in temperature and separation}
\label{Entropia_2_esferas_Drude}
\end{center}
\end{figure}

Once we have the asymptotic $\mathbb{T}$ matrix, obtaining the Casimir energy is a straightforward calculation. Carrying out the sum over Matsubara frequency, we obtain the Casimir contribution to the Helmholtz free energy as
\begin{align}
E_{T} & = -\frac{2 k_{B}T R^{6}e^{5 z}}{d^{6}\left(e^{2 z} - 1\right)^5}\times\nonumber\\
& \left(\left(6-10 z^2+22 z^4\right) \cosh(z) + \left(-36z+12z^3\right) \sinh(z) \right.\nonumber\\
& + \left(12 z+4z^3\right) \sinh(3 z) + 3 \cosh(5 z) \nonumber\\
& + \left.\left(-9 + 10z^{2} + 2z^{4}\right) \cosh(3 z) \right),
\end{align}
where $z = d/\lambda_{T}$. The zero- and high-temperature Casimir energy limits are given by
\begin{equation}
E_{0} = -\frac{23 \hbar c R^{6}}{4\pi d^{7}},
\hspace{20pt}
E_{cl} = - k_{B}T\frac{3 R^{6}}{d^{6}}.
\end{equation}
These asymptotic Casimir energies are universal because they do not depend on any material property. The lack of magnetic polarizability contributions is characteristic of the Drude model, leading to an effective scalar Dirichlet problem.
The entropy is obtained as in the perfect metal case. As only the electric polarization contributes, we expect the usual behavior of the entropy as a monotonic function of $\frac{d}{\lambda_{T}}$, as shown in Fig. \ref{Entropia_2_esferas_Drude}. The zero- and high-temperature Casimir entropy limits are given by
\begin{equation}
S_{0} = \frac{704 k_{B}R^{6}}{315 d}\left(\frac{\pi k_{B}T}{\hbar c}\right)^{5} - \frac{11776 k_{B}d R^{6}}{1575 }\left(\frac{\pi k_{B}T}{\hbar c}\right)^{7},
\end{equation}
\begin{equation}
S_{cl} = 3k_{B}\frac{R^{6}}{d^{6}}.
\end{equation}
In this case, the correction term to the quantum limit tends to reduce the entropy, but it is not strong enough to change the sign of it, as seen in Fig. \ref{Entropia_2_esferas_Drude}.

\section{Thermodynamical Consequences}\label{Sect:Thermodynamic}
In this section we discuss the thermodynamical consequences of the obtained results. In this article we have obtained and compared the large separation limit of the Casimir energy and entropy for two spheres using three different electric susceptibility models. For perfect metal spheres, at any nonzero fixed temperature, we observe an interval of separations for which entropy is negative, while at zero temperature the entropy is always zero.

In addition, the Casimir force is attractive for all separations and temperatures. So, we could naively think that we have possible violations of the third and second laws of thermodynamics, due to the existence of processes where the entropy of the system tends to increase and to the negative entropy intervals at finite temperature and distances, respectively.

According to the Krein formula \cite{Wirzba}, we know that the Helmholtz free energy of the electromagnetic field has three independent additive contributions (see Eq. \eqref{Definicion_DoS4} of the Appendix): one is from the thermal bath, being proportional to the volume of the space \cite{Spectra_Finite_Systems}. Another is the sum of contributions of objects immersed in the bath considered as isolated objects, each contribution being also a function of the volume and surface of each object \cite{Spectra_Finite_Systems}. The third is of geometrical nature, which we could call the Casimir part of the Helmholtz free energy and that we actually calculate in Eq. \eqref{Energy_T_finite}. This term depends on the electromagnetic nature of the objects in the system, but it has a geometric nature, because it depends on the relative separations and orientations between the objects, being zero iff the relative separations between all the considered objects become infinite.

Considering the whole system, the nonmonotonicity of the Casimir entropy with temperature is compensated by the contribution of the vacuum, because one scales with the global volume~\cite{Spectra_Finite_Systems} (Eq. \eqref{Definicion_DoS_Weyl}) and the other with the separation between the spheres (Eq. \eqref{High_T_Entropy}). This is not the case for the nonmonotonicity with separation. At constant temperature, the entropies of the thermal bath and of each object remain constant, but the Casimir force is always attractive while the entropy can increase or decrease with separation. Therefore, the potential violation of the second law of thermodynamics still requires an explanation.

The Krein formula also states that internal sources of entropy inside the spheres cannot appear in order to compensate these regions of anomalous behavior of the entropy, unless these internal sources are independent of the separation between the spheres.

The second law of thermodynamics states that global entropy must increase for any process, but only in closed systems. As we are working in the canonical ensemble, we are implicitly assuming that there exists an external reservoir which keeps our system at constant temperature, so the system is not isolated and entropy can increase or decrease without violation of the second law. In the canonical ensemble, the condition equivalent to the second law is that the global Helmholtz free energy must decrease for any process, and this is true for the studied system. Therefore, the appearance of nonmonotonic entropy behavior in the canonical ensemble just implies nonmonotonic behavior of the force with temperature, as seen in Eq. \eqref{Relacion_Fuerza_Entropia}.

\section{Conclusions}
In this article we have obtained and compared the large separation limit of the Casimir energy and entropy for two spheres using three different electric susceptibility models. For perfect metal spheres, at any nonzero fixed temperature, we observe an interval of separations for which entropy is negative, while at zero temperature the entropy is always zero.

We can trace the origin of this anomalous effect to the functional form of the Casimir energy in this large separation regime. It consists of four components, two of which are of attractive nature, equivalent to the sum of two scalar problems, one for the electric polarization and the other for the magnetic polarization. The other two components come from the cross-coupling between the electric and magnetic polarizations of the two spheres, and tend to reduce the Casimir energy between the spheres (Eq. 12 of \cite{EGJK} and Eq. 6 of \cite{PSA_EM_general}). These cross-coupling polarization terms are responsible for the impossibility of factorization of the electromagnetic Casimir energy into two equivalent scalar problems in general, and appear because the translation matrix $\mathbb{U}_{\alpha\beta}$ is not diagonal in polarization space. In this article we show that these cross terms not only reduce the Casimir energy, but also for some separations and temperatures, their contribution to the entropy is greater than the contributions of direct coupling between the electric and magnetic polarizabilities, resulting in an interval of negative entropy and nonmonotonic behavior of the entropy with separation and temperature.

In addition, the Casimir force is attractive for all separations and temperatures. So, we could naively think that we have possible violations of the third and second laws of thermodynamics, due to the existence of processes where the entropy of the system tends to increase and to the negative entropy intervals at finite temperature and distances, respectively. 

In Sect. \ref{Sect:Thermodynamic}, having into account the complete thermodynamical system, that the system is described by the canonical ensemble, and with the help of the Krein formula and Weyl formula, we have demonstrated that there are not violations of second and third laws respectively. Therefore, the appearance of nonmonotonic entropy behavior in the canonical ensemble just implies nonmonotonic behavior of the force with temperature, as seen in Eq. \eqref{Relacion_Fuerza_Entropia}.

The interval of negative entropy for the spheres appears because of the cross-coupling between the polarizations of the electromagnetic field, which leads us to conclude that it is a characteristic of the electromagnetic field and does not have an analog in the Casimir effect due to scalar fields.

Applying the PFA to this problem, we obtain that, for perfect metal spheres near contact, the entropy is always positive, so we performed a numerical study of the entropy between perfect metal spheres of equal radius at smaller separations. The results showed that there exists a minimum separation between the spheres for which the negative entropies disappear. In addition, the region of negative slope of the entropy with separation disappears for another smaller separation.

We have also obtained the energy and entropy for the plasma model, for which similar results are obtained for plasma penetration length $\lambda_{P} \lesssim 2R$, while for $\lambda_{P} > 2R$, the negative entropies disappear, and for another greater $\lambda_{P}$, the nonmonotonic behavior of the entropy with separation also disappears.

In the transparent limit ($\lambda_{P} \gg R$), the magnetic polarization does not contribute to the asymptotic Casimir effect, so the problem reduces to a scalar field problem and we find that entropy is a positive monotonic function.

When we study the asymptotic Casimir effect with Drude model spheres, the system is qualitatively different. The magnetic polarization does not contribute to the asymptotic Casimir effect because it depends on $d^{-4}$ instead of the $d^{-3}$ dependence of the electric polarization term. Then also for Drude spheres the asymptotic limit reduces to a scalar field problem, resulting in a positive monotonic entropy.

Nonmonotonicities of the Casimir force between objects are not unusual. As the Casimir effect is not pairwise additive, the interaction between two objects is affected by the presence of a third, leading to nonmonotonicities of the Casimir force between cylinders \cite{Fuerza_cilindros_en_presencia_de_plato,Fuerza_cilindros_en_presencia_de_plato2}, or between spheres \cite{Fuerza_esferas_en_presencia_de_plato} in the presence of a plate. The nonmonotonicity presented in this article has a different nature, as it does not come from interactions with a third object but rather from correlations with temperature, as seen in Eq. \eqref{Relacion_Fuerza_Entropia}.

In addition, in the assumptions of the multiscattering formalism it is implicitly assumed that the objects are stationary, at fixed positions in space, so all results shown here are valid for quasistatic processes. Therefore, we must carefully consider when this quasistatic assumption does not apply, because the stationary system could abandon equilibrium, requiring more careful study \cite{Intravaia_non_eq_and_dynamic_Casimir_effects,Dedkov_Kyasov}.

\acknowledgments
I acknowledge helpful discussions with R.~Brito, J.~Parrondo, J.~Horowitz, E.~Rold\'an, L.~Dinis, T.~Emig, and A.~Canaguier-Durand. This research was supported by the projects MOSAICO and MODELICO and an FPU MEC grant.

\section*{Appendix: Krein formula from multiscattering formalism}
In this appendix, we will derive the Krein formula of the electromagnetic field \cite{Wirzba} from the multiscattering formula \cite{EGJK}.

Krein formula factorizes the density of states of the electromagnetic field in three parts \cite{Wirzba}. Here we present a derivation of such formula from the Multiscattering formalism of the Casimir effect \cite{EGJK, Review-Jamal-Emig}.

The partition function ($\mathcal{Z} = {\sum_{n=0}^{\infty}}'\mathcal{Z}_{n}$) of the EM field in the presence of $N$ general dielectrics of arbitrary geometry is obtained, before regularization of the Casimir energy, as~\cite{Review-Jamal-Emig}
\begin{equation}\label{Z_Casimir_Completa}
\mathcal{Z}_{n} = \frac{1}{\Det{S_{n}}\Det{\mathcal{M}}},
\end{equation}
where $\Det{S_{n}}^{-1} = \Det{\Delta + \kappa_{n}^{2}}^{-1}$ is the $\text{n}^{\underline{\text{th}}}$ Matsubara frequency contribution to the partition function of the thermal bath, and $\mathcal{M}$ is the next $N\times N$ block matrix
\begin{equation}
\mathcal{M}_{\alpha\beta} = \delta_{\alpha\beta}\mathbb{T}_{\alpha}^{-1} + (\delta_{\alpha\beta} - 1)\mathbb{U}_{\alpha\beta}.
\end{equation}
If we multiply and divide Eq. \eqref{Z_Casimir_Completa} by $\Det{\mathcal{M}_{\infty}}$ (being $\mathcal{M}_{\infty}$ the same $\mathcal{M}$ matrix as above, but with each object at an infinite distance from each other), the Helmholtz free energy of the whole system can be writen as ($\beta\mathcal{F} = \log(\mathcal{Z})$)
\begin{equation}\label{E_Helmholtz_global}
\beta\mathcal{F} = - {\sum_{n=0}^{\infty}}'\left[\log\Det{S_{n}} + \log\Det{\mathcal{M}_{\infty}} + \log\left(\frac{\Det{\mathcal{M}}}{\Det{\mathcal{M}_{\infty}}}\right)\right],
\end{equation}
where $\log\Det{S_{n}}$ is the contribution of the thermal bath in absense of inmersed objects, which leads to the Planck formula of the blackbody spectrum, $\log\Det{\mathcal{M}_{\infty}} = \sum_{\alpha = 1}^{n}\log\Det{\mathbb{T}_{\alpha}}$ is the contribution to the Helmholtz free energy of each dielectric object immersed in the bath considered as an isolated object, and $\log\left(\frac{\Det{\mathcal{M}}}{\Det{\mathcal{M}_{\infty}}}\right) = \log\Det{\mathbb{I} - \mathbb{N}}$ is the Casimir part of the Helmholtz free energy, which depends on the electromagnetic nature of the objects in the system, but it has a geometric nature, because it depends on the relative separations and orientations between the objects, being zero iff the relative separations between all the considered objects become infinite. This term is the Casimir energy that we actually calculate in Eq. \eqref{Energy_T_finite}.

The energy of a quantum system is related with the Density of States $\rho(\omega)$ by~\cite{Wirzba}
\begin{equation}\label{Definicion_DoS}
E = \int_{0}^{\infty}d\omega\frac{\hbar}{2}\omega\rho(\omega).
\end{equation}
Using Eq. \eqref{E_Helmholtz_global}, the same energy in the zero temperature case can be writen as
\begin{equation}\label{E_Helmholtz_global1}
E = \frac{\hbar c}{2\pi}\int_{0}^{\infty}d\kappa\left[\log\Det{S_{\kappa}} - \sum_{\alpha = 1}^{n}\log\Det{\mathbb{T}_{\alpha}} + \log\Det{\mathbb{I} - \mathbb{N}}\right].
\end{equation}
Because the dispersion relation of the massless EM field $\omega = c\kappa$, and integrating by parts, we obtain
%\begin{equation}\label{E_Helmholtz_global2}E = \int_{0}^{\infty}d\omega\frac{\hbar}{2}\omega\frac{1}{\pi}\left[- \partial_{\omega}\log\Det{S_{0}} + \sum_{\alpha = 1}^{n}\partial_{\omega}\log\Det{\mathbb{T}_{\alpha}} - \partial_{\omega}\log\Det{\mathbb{I} - \mathbb{N}}\right].\end{equation}
\begin{align}\label{E_Helmholtz_global2}
E & = \int_{0}^{\infty}d\omega\frac{\hbar}{2}\omega\frac{1}{\pi}\times\\
& \times\left[- \partial_{\omega}\log\Det{S_{\omega}} + \sum_{\alpha = 1}^{n}\partial_{\omega}\log\Det{\mathbb{T}_{\alpha}} - \partial_{\omega}\log\Det{\mathbb{I} - \mathbb{N}}\right].\nonumber
\end{align}
Then, because Eqs. \eqref{Definicion_DoS} and \eqref{E_Helmholtz_global2}, we can relate the Density of States with scattering properties of the system as
\begin{equation}\label{Definicion_DoS2}
\rho(\omega) = \frac{1}{\pi}\left[- \partial_{\omega}\log\Det{S_{\omega}} + \sum_{\alpha = 1}^{n}\partial_{\omega}\log\Det{\mathbb{T}_{\alpha}} - \partial_{\omega}\log\Det{\mathbb{I} - \mathbb{N}}\right].
\end{equation}
As a conclussion, we can factorize the Density of States as the sum of three independent terms
\begin{equation}\label{Definicion_DoS4}
\rho(\omega) = \rho_{0}(\omega) + \sum_{\alpha = 1}^{n}\rho_{\alpha}(\omega) + \rho_{C}(\omega),
\end{equation}
where $\rho_{0}(\omega) = - \frac{1}{\pi}\partial_{\omega}\log\Det{S_{0}}$ is the Density of States of the free EM field, $\rho_{\alpha}(\omega) = \frac{1}{\pi}\partial_{\omega}\log\Det{\mathbb{T}_{\alpha}}$ is the change of the vacuum Density of States of the EM field because each $\alpha^{\text{\underline{th}}}$ dielectric object considered as a isolated object in the medium, and $\rho_{C}(\omega) = - \frac{1}{\pi}\partial_{\omega}\log\Det{\mathbb{I} - \mathbb{N}}$ is the change of the Density of States related with the relative positions of the $N$ dielectric objects, which is zero when all the objects are infinitely apart from each other. $\rho_{C}(\omega)$ is the source of the Casimir effect.

Eq. \eqref{Definicion_DoS4} is the Krein formula \cite{Wirzba}, derived from the Multiscattering formalism of the Casimir effect for the EM field. In \cite{Wirzba}, Eq. \eqref{Energy_T_finite} at zero temperature is obtained from the Krein formula removing the divergent contributions. Here we have followed the opposite way, a derivation of the density of states of the a thermal bath with intrusions and it factorization in Eq. \eqref{Definicion_DoS4}.

Terms $\rho_{0}(\omega)$ and $\rho_{\alpha}(\omega)$ diverges with $\omega$ because main contribution to Weyl formula for 3D volumes of the EM field~\cite{Spectra_Finite_Systems, Libro_Tejero}, 
\begin{equation}\label{Definicion_DoS_Weyl}
\rho_{V}(\omega) = \frac{V}{\pi^{2}c^{3}}\omega^{2}\theta(\omega),
\end{equation}
but $\rho_{C}(\omega)$ converges if our system consists on $N$ compact objects, because in this case, $\mathbb{N}$ can be demonstrated to be a trace class operator for all $\omega$, then the determinant is well defined~\cite{Casimir_forces_in_a_T-operator_approach}.

%\newpage
%\nocite*
%Para Bibtex, se usan las siguientes instrucciones:
%\bibliography{References}

\begin{thebibliography}{20}
\bibitem{Casimir Placas Paralelas} H.B.G. Casimir, Proc. K. Ned. Akad. Wet. 51, 793 (1948).
\bibitem{VdW int. electrica} H.B.G. Casimir and D. Polder, Phys. Rev. \textbf{73}, 360 (1948).
\bibitem{Schwinger}J. Schwinger, Lett. Math. Phys. \textbf{1}, 43 (1975).
\bibitem{EGJK}T. Emig, N. Graham, R.L. Jaffe, and M. Kardar, Phys. Rev. Lett. \textbf{99}, 170403 (2007). DOI: 10.1103/PhysRevLett.99.170403.
\bibitem{Review-Jamal-Emig}S.J. Rahi, T. Emig, N. Graham, R.L. Jaffe, and M. Kardar, Phys. Rev. D \textbf{80}, 085021 (2009). DOI: 10.1103/PhysRevD.80.085021.
\bibitem{Multiscattering_Lambrecht}A. Lambrecht P. A. Maia Neto and S. Reynaud. New Journal of Physics \textbf{8}, 243 (2006). DOI: 10.1088/1367-2630/8/10/243
% The Casimir effect within scattering theory
% Multiscattering desde formula placas cutre, pero bueno
\bibitem{Multiscattering_Milton}K. A. Milton and J. Wagner. J. Phys. A: Math. Theor. \textbf{41}, 155402 (2008). DOI: 10.1088/1751-8113/41/15/155402
% Multiple scattering methods in Casimir calculations
% Multiscattering sin Multiscattering, bueno tambien
\bibitem{Metodo_Caminos_Esfera_Placa_Escalar}A. Weber and H. Gies, Phys. Rev. Lett. \textbf{105}, 040403 (2010). DOI: 10.1103/PhysRevLett.105.040403.
% Nonmonotonic Thermal Casimir Force from Geometry-Temperature Interplay
% Con el metodo de caminos, calcula la no monotonicidad de la energia (thermal force) entre placa-esfera y plano-cilindro para el caso Dirichlet escalar.
\bibitem{Entropias_negativas_placas_Drude}V.B. Bezerra, G.L. Klimchitskaya, and V.M. Mostepanenko, Phys. Rev. A \textbf{65}, 052113 (2002). DOI: 10.1103/PhysRevA.65.052113
% Thermodynamical aspects of the Casimir force between real metals at nonzero temperature
\bibitem{Bordag_Entropia_Esfera_Placa}M. Bordag and I. G. Pirozhenko. Phys. Rev. D \textbf{82}, 125016 (2010). DOI: 10.1103/PhysRevD.82.125016
% Casimir entropy for a Drude ball in front of a Drude plane
\bibitem{Canaguier-Durand_Caso_Esfera_Placa_PRA}A. Canaguier-Durand, P.A. Maia Neto, A. Lambrecht, and S. Reynaud, Phys. Rev. A \textbf{82}, 012511 (2010). DOI: 10.1103/PhysRevA.82.012511.
% Thermal Casimir effect for Drude metals in the plane-sphere geometry, pero en PRA y mas completo
% Estudio Esfera-Placa, obtienen entropias negativas para caso Drude, para Plasma y para Metal Perfecto.
\bibitem{Wittmann}R.C. Wittmann, IEEE Trans. Antennas Propag. \textbf{36}, 8 (1988).
% Spherical Wave Operators and the Traslation Formaulas
% Matrices de traslacion esfericas
\bibitem{PSA_EM_general}P. Rodriguez-Lopez. Phys. Rev. E \textbf{80}, 061128 (2009). DOI: 10.1103/PhysRevE.80.061128
% Pairwise summation approximation of Casimir energy from first principles
\bibitem{Barton}G. Barton, Phys. Rev. A, \textbf{64}, 032102 (2001).
% Long-range Casimir-Polder-Feinberg-Sucher intermolecular potential at nonzero temperature
\bibitem{Ingold}G.-L. Ingold, A. Lambrecht, and S. Reynaud, Phys. Rev. E \textbf{80}, 041113 (2009). DOI: 10.1103/PhysRevE.80.041113.
% Quantum dissipative Brownian motion and the Casimir effect
% Intenta explicar la aparicion de entropias negativas a partir de la no conmutatividad de limites
\bibitem{Zandi_Emig_Placa_Esfera_varios_modelos}R. Zandi, T. Emig, and U. Mohideen. Phys. Rev. B \textbf{81}, 195423 (2010). DOI: 10.1103/PhysRevB.81.195423
% Quantum and thermal Casimir interaction between a sphere and a plate: Comparison of Drude and plasma models
% Da un review de las energías de Casimir entre esfera y placa para los modelos de metal perfecto, Plasma y Drude.
\bibitem{Wirzba}A. Wirzba, J. Phys. A: Math. Theor. \textbf{41} 164003 (2008). DOI: 10.1088/1751-8113/41/16/164003
% The Casimir effect as a scattering problem
% Aqui aparece la formula de Krein
\bibitem{Spectra_Finite_Systems} H.P. Baltes and E.R. Hilf, \textit{Spectra of Finite Systems}, (Birkh\"{a}user Boston, 1980).
% Libro en el que se da la expansión asintótica de la densidad de estados de recintos finitos, para los campos escalares y para el campo EM para recintos metales perfectos.

\bibitem{Fuerza_cilindros_en_presencia_de_plato}A. Rodriguez, M. Ibanescu, D. Iannuzzi, F. Capasso, J.D. Joannopoulos, and S.G. Johnson, Phys. Rev. Lett. \textbf{99}, 080401 (2007). DOI: 10.1103/PhysRevLett.99.080401
% Computation and Visualization of Casimir Forces in Arbitrary Geometries: Nonmonotonic Lateral-Wall Forces and the Failure of Proximity-Force Approximations
\bibitem{Fuerza_cilindros_en_presencia_de_plato2}S.J. Rahi, T. Emig, R.L. Jaffe, and M. Kardar, Phys. Rev.  A \textbf{78}, 012104 (2008). DOI: 10.1103/PhysRevA.78.012104
%Casimir forces between cylinders and plates
\bibitem{Fuerza_esferas_en_presencia_de_plato}P. Rodriguez-Lopez, S.J. Rahi, and T. Emig, Phys. Rev. A \textbf{80}, 022519 (2009). DOI:  10.1103/PhysRevA.80.022519
% Three-body Casimir effects and nonmonotonic forces
\bibitem{Intravaia_non_eq_and_dynamic_Casimir_effects}F. Intravaia, C. Henkel, and M. Antezza, in \textit{Casimir Physics}, edited by D. Dalvit, P. Milonni, D. Roberts, and F. da Rosa. Springer 2011. ISBN: 978-3-642-20287-2
% Fluctuation-induced forces between atoms and surfaces: the Casimir-Polder interaction
% Capitulo de un libro de pronta publicacion en el que se hace un review de Efecto Casimir fuera del equilibrio y dinamico
\bibitem{Dedkov_Kyasov}G.V. Dedkov and A.A. Kyasov, Nucl. Instrum. Methods Phys. Res. B 268, 599 (2010).
%Tangential force and heating rate of a neutral relativistic particle mediated by equilibrium background radiation



\bibitem{Libro_Tejero}M. Baus and Carlos F. Tejero, \textit{Equilibrium Statistical Physics: Phases of Matter and Phase Transitions,} (Springer, 2007). ISBN: 3540746315.

\bibitem{Casimir_forces_in_a_T-operator_approach}O. Kenneth and I. Klich. Phys. Rev. B \textbf{78}, 014103 (2008). DOI: 10.1103/PhysRevB.78.014103
% Casimir forces in a T-operator approach






%\bibitem{Bordag_Energia_Esfera_Placa}M. Bordag and I. G. Pirozhenko. Phys. Rev. D \textbf{81}, 085023 (2010). DOI: 10.1103/PhysRevD.81.085023
% Vacuum energy between a sphere and a plane at finite temperature
% Energía libre con PFA a partir de multiscattering, no entropias

\thispagestyle{empty}
\end{thebibliography}
%\bibliographystyle{unsrt}
% Para el caso que nos ocupa, lo haré de la forma cutre:

\end{document}